\documentclass[%
 reprint,              
 amsmath,amssymb,
 aps,
 prl,
 floatfix
]{revtex4-2}

\usepackage{graphicx}        
\usepackage{svg}             
\usepackage{hyperref}        

\usepackage{bm}              
\usepackage{physics}         
\usepackage{xcolor}          

\usepackage[normalem]{ulem} 

\usepackage[sort&compress]{natbib}  

\setsvg{inkscapelatex=false}  

\begin{document}

\title{Snap-Through Symmetry Breaking in Multicellular Aggregates}
\author{Mart\'i Planasdemunt-Hospital$^{1,2,3}$}
\email{marti.planasdemunt@upc.edu}

\affiliation{$^{1}$Department of Physics, Universitat Politècnica de Catalunya - BarcelonaTech (UPC), 08028, Barcelona, Spain}
\affiliation{$^{2}$Institute for Research and Innovation in Health (IRIS), Universitat Politècnica de Catalunya - BarcelonaTech (UPC), 08028, Barcelona, Spain}
\affiliation{$^{3}$Barcelona Collaboratorium for Modelling and Predictive Biology, 08003, Barcelona, Spain}

\author{David Oriola$^{1,2,3}$}
\affiliation{$^{1}$Department of Physics, Universitat Politècnica de Catalunya - BarcelonaTech (UPC), 08028, Barcelona, Spain}
\affiliation{$^{2}$Institute for Research and Innovation in Health (IRIS), Universitat Politècnica de Catalunya - BarcelonaTech (UPC), 08028, Barcelona, Spain}
\affiliation{$^{3}$Barcelona Collaboratorium for Modelling and Predictive Biology, 08003, Barcelona, Spain}

\date{\today}

\begin{abstract}
Symmetry breaking is a fundamental process in biological organization. Yet the physical mechanisms that control its onset in multicellular systems remain poorly understood. Motivated by gastruloids -- multicellular aggregates in which cells differentiate and spontaneously break symmetry --  we study a two-phase viscoelastic solid drop as a minimal model of tissue polarization. We show that elastocapillary competition fundamentally alters the classical engulfing transition of fluid drops. While symmetry breaking occurs continuously in the purely viscous limit, bulk elasticity instead leads to a discontinuous snap-like transition. We demonstrate that this behavior can be understood as a wetting transition governed by a spreading coefficient which depends on the drop size and phase volume fraction. Applying our model to gastruloids, we find a critical volume fraction for symmetry breaking, providing a self-organized mechanism for robust developmental timing. Our results identify a generic elastocapillary transition that enables spontaneous symmetry breaking in biphasic drops.
\end{abstract}

\maketitle

Symmetry-breaking processes are ubiquitous in biology across multiple scales~\cite{li2010symmetry,goehring2011polarization,motegi2013network,zhang2018symmetry,ierushalmi2021cytoskeletal}. At the cellular level, examples include cell polarization~\cite{mori2008wave,munro2009cellular,rappel2017mechanisms}, apical--basal polarity establishment~\cite{nelson2009remodeling,vladar2009planar}, cell division~\cite{chang2009shaping,slaughter2009symmetry}, and axon polarization~\cite{tahirovic2009neuronal}. Such processes often rely on mechanochemical amplification mechanisms capable of converting transient local fluctuations or noise into robust system-scale asymmetries~\cite{goehring2011polarization,motegi2013network,ierushalmi2021cytoskeletal}. By contrast, symmetry breaking at the supracellular scale remains less understood~\cite{ishihara2018spontaneous}. In multicellular systems, symmetry breaking emerges from cell--cell interactions mediated by signalling and mechanics. During embryonic development, spatial symmetries must be broken in a reproducible manner in order to form the animal body axes~\cite{beddington1999axis,benazeraf2013formation,anlas2021studying}. In embryos, symmetries are often broken as a consequence of external signals coming from extraembryonic tissues~\cite{sozen2021dynamics}. However, in some cases, such as in mammalian blastocysts~\cite{wennekamp2013self,bedzhov2014developmental}, radial symmetry breaking is spontaneous. Moreover, in development, the timing of symmetry breaking must be tightly coordinated with cellular differentiation. This requirement suggests that tissues need to dynamically tune their mechanical and biochemical properties in order to cross bifurcation points that trigger morphogenetic transitions. How evolving cell populations couple to tissue mechanics to robustly trigger symmetry breaking remains unclear. \\

\begin{figure}[h!]
    \centering
    \includegraphics[width=\linewidth]{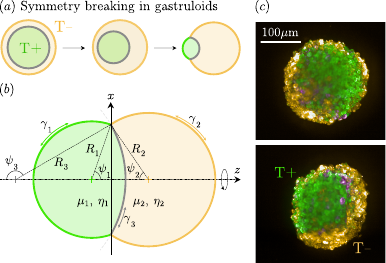}
    \caption{Gastruloids as two-phase drops. (a) Symmetry breaking in gastruloids; arrows indicate time evolution. (b) Parametrization of the model with coordinates $R_i$, $\psi_i$ and mechanical parameters $\gamma_i$, $\mu_k$ and $\eta_k$ for $i=1,2,3$ and $k=1,2$. (c) Confocal fluorescence microscopy images of gastruloids, extracted from Ref.~\cite{oriola2024cell}. T/Brachyury gene expression (T+) is labeled in green and Gata6/Adlh1a2 gene expression (T-) in yellow.}
    \label{fig:1}
\end{figure}

In the last decade, simplified 3D reconstituted systems have been shown to break symmetry {\it in vitro} in the absence of external asymmetric cues~\cite{ten2008wnt,van2014symmetry,meinhardt20143d,takata2017self}. From all the previous systems, 3D {\it gastruloids} have received great attention as a minimal system to study spontaneous symmetry breaking~\cite{hashmi2022cell,oriola2024cell,sozen2021dynamics,mcnamara2024recording,gsell2025marangoni}. These structures recapitulate anteroposterior axis formation {\it in vitro} through the polarization of a group of cells expressing the gene {\it T/Brachyury} (hereinafter referred to as T)~\cite{van2014symmetry,beccari2018multi}. Several studies support cell sorting as the primary mechanism driving phase separation of T-expressing cells (T+) from non-expressing, T-negative cells (T-)~\cite{mcnamara2024recording,oriola2024cell,gsell2025marangoni}. However, it is still unclear how the T+ phase polarizes, thus defining a preferred axis in the system. Previous work has shown the existence of a transient T+ phase enveloped by T- cells prior to polarization~\cite{oriola2024cell} (see Fig.~\ref{fig:1}a,c), reminiscent of the patterns observed in classic {\it in vitro} cell sorting studies~\cite{foty1996surface,foty2005differential,schotz2008quantitative}. Motivated by this physical picture, here we regard gastruloids as two-phase viscoelastic solid drops. The elastic nature of the two phases is motivated by previous work where we showed that T+ and T- cellular aggregates behave as viscoelastic solid drops when fused homotypically~\cite{oriola2022arrested,oriola2024cell}. The two-phase picture has been used before to model gastruloids~\cite{gsell2022phase,gsell2025marangoni}, and it is further supported by the fact that merging T+ and T- aggregates leads to gastruloid-like structures~\cite{xu2021construction}.  Here, we find that the presence of bulk elasticity drives a {\it snap-like} transition that breaks radial symmetry in the system. This is in contrast to the purely fluid case, where a continuous transition is observed~\cite{maitre2016asymmetric}. We show that polarization can be understood as a dewetting process with a modified spreading coefficient that depends on bulk elasticity. Finally, by using experimentally measured mechanical parameters for gastruloids, we find that the relative abundance of the two cell populations, expressed as a volume fraction, plays the role of a control parameter driving symmetry breaking. This provides a self-organized mechanism by which the next morphogenetic stage is only reached if the right cell proportions are obtained while keeping the tissue mechanical properties constant over time. \\

\textit{Theoretical Model}: We consider gastruloids as two-phase drops composed of two viscoelastic solid phases ($k=1,2$), one corresponding to cells expressing the T/Brachyury gene (phase 1, T+) and the other to cells not expressing the gene (phase 2, T-) (see Fig.~\ref{fig:1}a,b). We define the volume fraction $\varphi$ as the fraction of T- cells. For simplicity, we neglect cell proliferation and assume a constant total volume of the two-phase drop.  Here we follow the parametrization used in Ref.~\cite{maitre2016asymmetric} for the study of cell doublets. We prescribe a simplified geometry of the system as a combination of three spherical cap surfaces ($i=1,2,3$) whose centers are aligned along the $z$-axis, as shown in Fig.~\ref{fig:1}b. The external surfaces (1 and 2) separate the phases from the surrounding media, and the interface (3) separates the two phases. Motivated by previous experimental work~\cite{oriola2022arrested,oriola2024cell}, we model the bulk mechanics of the phases as Kelvin-Voigt materials with shear elastic moduli $\mu_k$ and shear viscosities $\eta_k$, for each phase $k=1,2$. Additionally, we consider effective tissue surface tensions $\gamma_i$ for each one of the three surfaces. A Rayleighian functional $\mathcal{R}=\dot{\mathcal{F}}+\mathcal{D}$~\cite{onsager1931reciprocal1,onsager1931reciprocal2,doi2011onsager,arroyo2017onsager} can be defined for the system, where $\mathcal{F}$ and $\mathcal{D}$ correspond to the free energy and dissipation functions, respectively, and read:
\begin{eqnarray}
\mathcal{F}&=&\sum_{k=1,2} \mu_k \int_{\Omega_k}  \, \boldsymbol{\varepsilon}_k:\boldsymbol{\varepsilon}_k \, dV+\sum_{i=1,2,3} \int_{\partial \Omega_i}\gamma_i\, dS \\
\mathcal{D}&=& \sum_{k=1,2} \eta_k \int_{\Omega_k}   \, \dot{\boldsymbol{\varepsilon}}_k:\dot{\boldsymbol{\varepsilon}}_k   \, dV
\end{eqnarray}
where $\boldsymbol{\varepsilon}_k$ and $\Omega_k$ are the symmetric strain tensor and bulk domain of integration for phase $k$, and  $\partial\Omega_i$ is the domain of integration of surface $i$. To simplify the problem, we assume a homogeneous biaxial extensional flow when the two phases evolve along the $z$-direction~\cite{bellehumeur1998role,kosztin2012colloquium,oriola2022arrested,samatas2024arrested}, such that $\boldsymbol{\varepsilon}_k=-\varepsilon_k \,  {\rm diag}(-1/2,-1/2,1)$, where $\varepsilon_k$ depends on the geometry of the problem (see Supplemental Material~\cite{SupplementalMaterial}). Finally, we introduce the Lagrange multipliers $p_1$ and $p_2$ to enforce a volume constraint for the two phases, forming the constrained Lagrangian  $\mathcal{L}=\mathcal{R}-\sum_{k=1,2} p_k \mathcal{Q}_k$, where $\mathcal{Q}_k=\dot{V}_k-\dot{V}_{0,k}$, with $V_k$ the volume of phase $k$ and $V_{0,k}=\frac{4\pi}{3}R_{0,k}^3$. Here, $R_{0,1}=R_02^{1/3}\left(1-\varphi\right)^{1/3}$ and $R_{0,2}=R_02^{1/3}\varphi^{1/3}$ such that the total volume is conserved $V_{0,1}+V_{0,2}=8\pi R_0^3/3$, and $R_0$ corresponds to the radius of the two aggregates for $\varphi=1/2$ when the contact area is zero ($\psi_1=\psi_2=0$). The dynamical equations of the system are found by minimizing the constrained Lagrangian with respect to the fluxes $\boldsymbol{x}'$ and the Lagrange multipliers, and follow the linear system of equations:
\begin{align}
    \mathbf{S}\begin{bmatrix}
        \boldsymbol{x}'\\
        \boldsymbol{p}
         \end{bmatrix}=\mathbf{b},\label{eq:system}
\end{align}
where $\mathbf{S}$ is a symmetric matrix, $\mathbf{b}$ is a vector defining the equilibrium configuration, and the Lagrange multipliers $\boldsymbol{p}$ play the role of the dimensionless pressures for each phase (see Supplemental Material~\cite{SupplementalMaterial}). The symbol ($'$) stands for a dimensionless time derivative $\text{d}/\text{d}\tau=\tau \text{d}/\text{d}t$, where $\tau=\eta_1 R_0/\gamma_1$ is the viscocapillary relaxation time of phase 1. The equilibrium configuration of the system can be found by setting $\boldsymbol{x}'=\boldsymbol{0}$, or equivalently, $\mathbf{b}=\mathbf{0}$ (see Supplemental Material~\cite{SupplementalMaterial}). 
In the purely fluid case ($\mu_k=0$), Young-Dupr\'e's equilibrium condition is recovered~\cite{maitre2016asymmetric}. However, in the presence of bulk elasticity ($\mu_k\neq0$), Young-Dupr\'e's relationship is modified, and it becomes dependent on the size of the system $R_0$ and the volume fraction $\varphi$ (see Supplemental Material~\cite{SupplementalMaterial}). Next, we will see that this result has crucial consequences for symmetry breaking. \\

\begin{figure*}
    \centering
    \includegraphics[width=\linewidth]{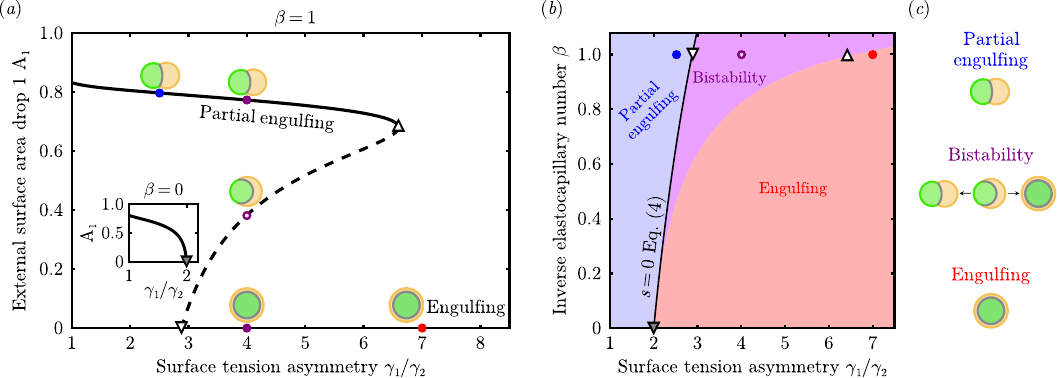}
    \caption{Snap-like engulfing transition in viscoelastic solid drops. (a) Bifurcation diagram with ($\beta=1$, solid case) and without (inset; $\beta=0$, fluid case) the presence of elasticity. Solid and dashed lines correspond to stable and unstable equilibrium branches, respectively. Filled and open bullets, with their adjacent schemes, denote stable and unstable equilibrium points, respectively. Upward triangles indicate saddle-node bifurcation points, and downward triangles indicate the limit equilibrium configuration with the lowest, non-zero $\mathrm{A}_1$ value computed. A similar bifurcation diagram is obtained if $\gamma_1/\gamma_3$ is varied by keeping $\gamma_1/\gamma_2$ constant and different from $1$ (see Fig.~\ref{fig:S1}). (b) A $\beta$ — $\gamma_1/\gamma_2$ phase diagram uncovers three regimes according to the degree of engulfing of the stable configuration. Horizontal cuts ($\beta=0$ and $\beta=1$) correspond to the bifurcation diagrams in (a). (c) Schematic description of the three possible regimes: partial engulfing, bistability and engulfing. The remaining parameter values are $\gamma_1/\gamma_3=2$ and $\varphi=1/2$.}
    \label{fig:2}
\end{figure*}

{\it Snap-like symmetry breaking}:
It is well known that in the purely fluid case, changes in the relative values of the different surface tensions ($\gamma_1,\gamma_2$ and $\gamma_3$) can induce an engulfment transition between the two phases~\cite{brodland2002differential,guzowski2012structure,maitre2016asymmetric}. Next, we assess how the presence of finite bulk elasticity ($\mu_k\neq 0$) affects this process. For simplicity, we will consider the case $\mu_1=\mu_2=\mu$ and define the dimensionless number $\beta=\mu R_0/\gamma_1$, which corresponds to an inverse elastocapillary number~\cite{oriola2022arrested,planasdemunt2026cell}. As an observable for engulfment, we define the normalized area of the external surface with the highest surface tension, which we denote (considering that $\gamma_1>\gamma_2$ in Fig.~\ref{fig:1}b) $\mathrm{A}_1=\hat{R}_1^2(1+\cos{\psi_1})/\left(2^{5/3}(1-\varphi)^{2/3}\right)$, with $\hat{R}_1\equiv R_1/R_0$~\cite{maitre2016asymmetric}. Using this definition, {\it engulfing} corresponds to $\mathrm{A}_1=0$, while \textit{partial engulfing} is found in the range $0<\mathrm{A}_1<1$~\cite{guzowski2012structure,maitre2016asymmetric}.  \\

\begin{figure}
    \centering
    \includegraphics[width=\linewidth]{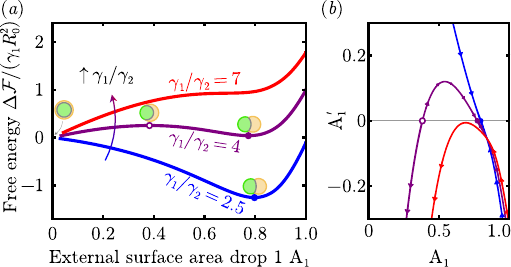}
    \caption{Spontaneous symmetry breaking. (a) As surface tension asymmetry $\gamma_1/\gamma_2$ increases, the free-energy landscape progressively flattens, revealing different equilibrium points (stable or unstable, shown with filled and open bullets, with their corresponding schemes). The difference $\Delta \mathcal{F}=\mathcal{F}-\mathcal{F}(\mathrm{A}_1=0)$ is plotted for convenience. $\mathcal{F}(\mathrm{A}_1=0)$ is obtained by interpolation. (b) The dynamics slow down near equilibrium points. Arrows indicate the direction of time evolution. The colors and $\gamma_1/\gamma_2$ values match those of the equilibrium points in Fig.~\ref{fig:2}.}
    \label{fig:3}
\end{figure}

For a given value of the interfacial surface tension $\gamma_3$, the surface tension asymmetry $\gamma_1/\gamma_2$ is known to
control a continuous engulfing transition in the purely fluid case ($\beta=0$)~\cite{maitre2016asymmetric}, as the interface curves (Fig.~\ref{fig:2}a inset). 
Notably, we find that bulk elasticity $(\beta \neq 0)$ gives rise to a saddle-node bifurcation (Fig.~\ref{fig:2}a), revealing a discontinuous engulfing transition. Stable and unstable equilibrium points are obtained by means of a continuation method (see Supplemental Material~\cite{SupplementalMaterial}). The unstable branch gives rise to a bistable region where both complete and partial engulfing configurations coexist. For a given value of the surface tension asymmetry $\gamma_1/\gamma_2$, an initial condition evolves toward partial engulfing if its $\mathrm{A}_1$ value exceeds that of the unstable equilibrium, and towards engulfing otherwise (see Movie 1 and Movie 2 of the Supplemental Material, respectively). Outside the bistable region, partial engulfing is stable for low tension asymmetry, whereas complete engulfing is stable at high tension asymmetry values. Starting from the stable branch, increasing $\gamma_1/\gamma_2$ beyond a critical value (upward triangle in Fig.~\ref{fig:2}a) triggers a snap-like transition to engulfment ($\mathrm{A}_1=0$). Elasticity can also lead to hysteresis: upon subsequently decreasing $\gamma_1/\gamma_2$, the system remains engulfed until it reaches the unstable branch (downward triangle in Fig.~\ref{fig:2}a). At this point, bistability is lost, and the system abruptly returns to the stable branch. In Fig.~\ref{fig:2}b, a phase diagram summarizes the three possible regimes. When bulk elastic forces are comparable to or larger than surface tension forces ($\beta \gtrsim 1)$ partial engulfing and bistability are favored against engulfing. Intuitively, elastic stresses oppose engulfment, as this process requires substantial deformation of the bulk of the enveloping phase. Finally, from an energetic perspective, the free-energy landscape resembles a first-order phase transition in Landau theory~\cite{landau1937theorie}, with $\mathrm{A}_1$ playing the role of the order parameter and $\gamma_1/\gamma_2$ acting as the control parameter (see Fig.~\ref{fig:3}a). In this picture, low $\gamma_1/\gamma_2$ values yield one stable asymmetric state, i.e., with $\mathrm{A}_1 \neq 0$. At higher values, an unstable equilibrium point emerges, which further on annihilates the former stable state (Fig.~\ref{fig:3}b). \\

{\it Volume fraction controls wetting}:  Engulfment can be seen as a particular form of wetting in which one drop (hereafter referred to as phase) spreads over another deformable drop, rather than over a rigid substrate~\cite{guzowski2012structure}. The spreading criterion is traditionally predicted by the \textit{spreading coefficient}~\cite{bonn2009wetting,douezan2011spreading,guzowski2012structure,douezan2012dewetting,perez2019active}. 
Considering that bulk elastic forces compete with surface tensions, we propose the following modification for the spreading coefficient based on dimensional analysis: 
\begin{equation}
    s=\gamma_1-\gamma_2-\gamma_3-a(\varphi)R_0\mu.\label{eq:s}
\end{equation}
 
where $a(\varphi)$ is an unknown dimensionless parameter which depends on the geometry of the problem and the volume fraction $\varphi$. At the engulfment transition ($s=0$, see Fig.~\ref{fig:2}b), Eq.~\eqref{eq:s} reduces to the linear relation $\gamma_2/\gamma_1=1-\gamma_3/\gamma_1-a(\varphi)\beta$, which we verify numerically (see Supplemental Material~\cite{SupplementalMaterial}), with $a(\varphi)$ following a polynomial dependence on $\varphi$ (see Fig.~\ref{fig:S2}). Thus, we find that bulk elasticity introduces the volume fraction $\varphi$ and aggregate size $R_0$ as control parameters for the engulfing transition, in contrast to the purely fluid case, where the transition is only controlled via surface tension differences. This result is key for tissues composed of different cell types, where cell fractions $\varphi$ are controlled through differentiation kinetics. \\

\begin{figure}
    \centering
    \includegraphics[width=\linewidth]{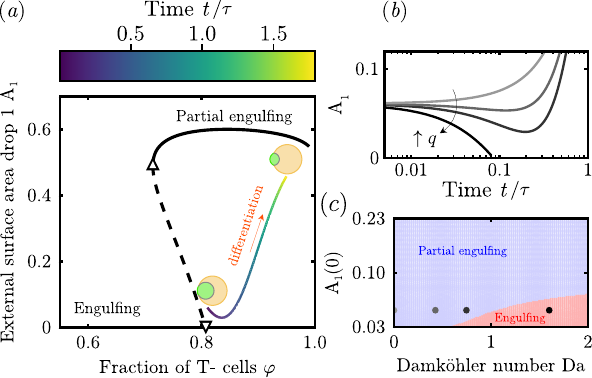}
    \caption{Cell-proportion control of symmetry breaking in gastruloids. (a) Bifurcation diagram using experimentally inferred parameters~\cite{oriola2024cell} (see Supplemental Material~\cite{SupplementalMaterial}).  Solid and dashed lines correspond to stable and unstable equilibrium branches, respectively. The colored line corresponds to a trajectory of the dynamical system with cell differentiation ($q=0.07 \, \text{h}^{-1}$, $\text{Da}\approx0.747)$. (b) Trajectories over time for increasing values of $q$ ($0, 0.04, 0.07, 0.15 \, \text{h}^{-1}$) starting at the same initial condition. Increasing the differentiation rate leads to engulfment.  (c) Phase diagram showing the interplay between the initial exposed area $\mathrm{A}_1(0)$ and the Damk\"ohler number $\text{Da}$. Each dot corresponds to a time trajectory in (b) with which it shares the shade of grey.}
    \label{fig:4}
\end{figure}

{\it Cell differentiation control of symmetry breaking}. We have shown that, unlike liquid droplets, engulfment in viscoelastic solid drops depends on the phase fraction $\varphi$, which means that morphological transitions can be regulated by tuning phase fractions without altering material properties. Such a strategy is particularly attractive in biological systems during development, where controlling cell population proportions is likely more accessible than dynamically tuning cellular mechanics. Cells can therefore retain fate-specific mechanical properties, while changes in their proportions alone drive large-scale morphological transitions. In this section, we exploit this result by applying our model to gastruloids, using the mechanical parameters obtained from experimental measurements~\cite{oriola2022arrested,oriola2024cell,planasdemunt2027} (see Supplemental Material). In gastruloids, a core of T+ cells is engulfed by T- cells with $\varphi \simeq 0.6$ at 48 hpa (hours post-aggregation). Later, T+ cells transition to T- cells up to $\varphi \simeq 0.8$ at 72 hpa~\cite{oriola2024cell}. In Fig.~\ref{fig:4}a, we show the bifurcation diagram that results from varying $\varphi$. Notably, unlike for liquid drops~\cite{maitre2016asymmetric} (Fig.~\ref{fig:S3}), the model predicts a saddle-node bifurcation around $\varphi \simeq 0.7$, which agrees with symmetry breaking in experimental observations. Next, instead of a constant volume fraction, we now allow for a time-varying fraction $\varphi(t)$ that evolves according to cell differentiation. In particular, in this time period $\varphi(t)$ was shown to follow simple first-order kinetics with $\dot{\varphi}=q(1-\varphi)$~\cite{oriola2024cell}, yielding the dynamics shown with a colored line in Fig.~\ref{fig:4} for that given initial condition. To compare the differentiation and the viscocapillary relaxation timescales, we define the corresponding Damk\"ohler number as ${\rm Da}=q \, \tau$. Interestingly, for the case of gastruloids we find ${\rm Da}\simeq 1$~\cite{oriola2024cell}. Hence, symmetry breaking in this case can be studied in the framework of {\it dynamical bifurcations}, where the control parameter evolves slowly over time~\cite{benoit2006dynamic,liu2021delayed}. Indeed, we observe delayed symmetry breaking (Fig.~\ref{fig:4}a, Movie 3 of the Supplemental Material), where the system is transiently trapped near the engulfing (unstable) state before finally breaking away to the partially engulfing (stable) state. However, cell differentiation not only can affect the dynamics of the system but also modify the final state (Fig.~\ref{fig:4}c). We find that whether the system can break away and polarize depends both on the Damk\"ohler number ${\rm Da}$ and the initial surface area of the T+ cell core exposed to the medium $\mathrm{A}_1(0)$, as shown in Fig.~\ref{fig:4}b,c. 
Polarization is abolished for high differentiation rates (${\rm Da}\gg1)$, while a large initial exposed area $\mathrm{A}_1(0)$ favors it. Thus we find that a fine balance between viscocapillary and differentiation timescales is required for symmetry breaking.  \\

{\it Discussion}. In this work, we have studied a symmetry-breaking process in a two-phase drop composed of viscoelastic solid phases. Symmetry breaking is understood here as a transition from an engulfing configuration, where one phase is inside the other, to the expulsion of the inner phase, defining an axis along their centers. While such a transition is continuous in liquid drops as a function of surface tension asymmetry, we find that the transition becomes discontinuous in the case of viscoelastic solid drops, leading to snap-like behavior. Elasticity favors partial engulfing and generates a bistable regime where complete and partial engulfing phases coexist. Most importantly, we show that bulk elasticity endows the system with a new control parameter for symmetry breaking, which is the volume fraction $\varphi$. This fact allows tissues to control phase transitions in development solely by dynamically controlling cell fate proportions. We test this hypothesis by applying our model to gastruloids and find a critical cell fraction for symmetry breaking, which is consistent with experimental evidence. \\

Given that gastruloids are active systems operating far from thermodynamic equilibrium, our approach should be regarded as an effective description based on emergent mechanical properties. Whether the observed morphologies in experiments correspond to effective mechanical equilibria or to long-lived transients is difficult to discern. In particular, one could argue that for a purely fluid two-phase drop, the apparent de-engulfing transition corresponds to increasingly slow relaxation dynamics rather than a genuine change in equilibrium morphology~\cite{gsell2022phase}. Then, partially engulfed configurations would correspond to long-lived transients~\cite{tordoff2021incomplete} that appear stable within the experimental observation window, while complete engulfment would remain the ultimate equilibrium configuration. Distinguishing between these scenarios requires experiments probing longer timescales. Nevertheless, our framework suggests that effective equilibrium arguments can capture dominant mechanisms underlying symmetry breaking in these systems. \\

We find that, in the presence of cell differentiation, symmetry breaking depends on the initial exposed area $\mathrm{A}_1(0)$ of the T+ phase to the medium (Fig.~\ref{fig:4}c,b). Although our model does not capture the processes that establish this initial configuration, we interpret the exposed area as an indicator of the active fluctuations that drive the T+ phase outwards. Starting from a fully engulfed configuration, small fluctuations may lead to a transient exposure of the T+ phase, which, according to our model, triggers symmetry breaking as long as the differentiation rate is slow enough. Agent-based simulations can shed light on how this process occurs at the cellular scale~\cite{oriola2024cell}. Finally, we also point out that the mechanical parameters assigned to the T- population in our model correspond to those measured for pluripotent cell aggregates~\cite{oriola2024cell}. This is an approximation, since the T- population during symmetry breaking is predominantly mesodermal, for which mechanical measurements were not available.\\

More generally, we uncover a dynamical bifurcation arising from the competition between elastic and capillary forces. Similar transitions have been observed in a variety of soft-matter systems, including elastic capsules~\cite{py2007capillary,roman2010elasto}, filaments~\cite{evans2013elastocapillary}, and adhesive contacts~\cite{style2013universal,headley2025elastocapillary}. Such phenomena are broadly known as snap transitions~\cite{liu2021delayed,fargette2014elastocapillary}. In this context, our results reveal that similar physics can govern the organization of multicellular systems, providing a generic route to symmetry breaking and polarization. \\

{\it Acknowledgements}. We thank R. Alert, S. Gsell, E.M. Hahn, F. J\"ulicher, M. Merkel, A. Torres-S\'anchez and H. Turlier for insightful discussions. We also thank the Barcelona Collaboratorium for its support through seminars and fruitful scientific interactions. D.O. and M.P.-H. acknowledge funding from the Spanish Ministry of Science, Innovation and Universities under project PID2024-155925NA-I00. M.P-H. also acknowledges the Joan Oró and Formació d'Investigadors Strategic Technologies for Europe Platform (FI STEP) grants supported by the Departament de Recerca i Universitats de la Generalitat de Catalunya and co-financed by the European Social Fund Plus (No. 2025 FI-1 00056 and No. 2025 STEP 00073).\\

{\it Data Availability}. The code used to generate the figures and reproduce the numerical results is at the moment available upon request at https://github.com/mplanasdemunt/snap-through-symmetry-breaking-in-multicellular-aggregates.git.


\bibliographystyle{apsrev4-2}
\bibliography{references}  

\end{document}